quant-ph/0407196

\documentclass[pra,12pt,eqsecnum]{revtex4}
\usepackage{graphicx}

\begin{document}

\def\BE{\begin{equation}}
\def\EE{\end{equation}}
\def\BY{\begin{eqnarray}}
\def\EY{\end{eqnarray}}

\def\L{\label}
\def\nn{\nonumber}
\def\({\left (}
\def\){\right)}
\def\[{\left [}
\def\]{\right]}
\def\o{\overline}
\def\i{\imath}

\title{Polarization statistical properties of the
emission from the single mode Vertical-Cavity Surface-Emitting Lasers with the equally
living laser levels}
\author{ Yu. M. Golubev and T. Yu. Golubeva }
\email{yugolubev@peterlink.ru, zernova@peterlink.ru}
\affiliation{V. A. Fock Physics Institute, St. Petersburg State University, ul.
Ul'anovskaya 1, 198504 St. Petersburg, Stary Petershof, Russia \\}
\author{ E. Giacobino }
\email{  elg@spectro.jussieu.fr}
\affiliation{ Laboratoire Kastler Brossel,
Universite Pierre et Marie Curie, F-75252 Paris Cedex 05, France\\}

\date{\today}

\begin{abstract}
The full quantum-statistical theory of the Vertical-Cavity Surface-Emitting
Laser (VCSEL) in the form of the Langevin equations is constructed for arbitrary
relations between the frequency parameters. The same theoretical treatment as in
Ref.~[1,2] are used. For detailed analysis the theory is applied for lasers with
equally living laser levels and on this basis the analytical expressions for the
spectral densities of the Stokes parameter fluctuations are obtained in the
explicit dependence on the physical phenomena, including the spin-flip and the
optical anisotropy. It is demonstrated the arbitrary distribution
 of electrons between the sub-levels under pumping does not restrict a possibility to achieve
 the noise reduction
  below the quantum limit. Under comparison with phenomenological
 treatment Ref.~\cite{3} it is shown  this approach turns out to be not quite satisfied.

\end{abstract}
\maketitle

\section{Introduction}\L{I}
In the last years there has been an increased interest to the polarization
properties of the VCSELs. This interest is motivated in the first line by the
potential applications of this type of lasers in the high-rate optical
communications Ref.~\cite{4}. But there is also more fundamental reason for
understanding of the polarization behavior in VCSELs, namely, a possibility of
generating the intensity-squeezed light using the sub-Poissonian pump of the
active medium Ref.~\cite{5,6}. To date, squeezing in VCSELs has been
demonstrated experimentally for both the single-mode operation and in the
multi-transverse-mode regime Ref.~\cite{7}.

Now in the different scientific groups  two models of laser are discussed in the
main. First of them is with the shortly living upper level and the other - with
the equaled lifetimes for both the levels. The known spin-flip theory \cite{8}
was elaborated for the first time  just for the latter system. Nevertheless a
lot of it is suitable for the first one too.

As for statistical aspects   the laser with the equally living levels was
studied only phenomenologically Ref.~\cite{3} in distinguish from the other
system which has been studied in details within the limits of the quantum
electrodynamics Ref.~\cite{1,2}. Our main goal here to develop not  less
qualitative theory for the system with the equally living levels. This will
allow us not only to write the different correct analytical expressions, but
also to estimate the possibilities of the phenomenological treatment
Ref.~\cite{3}.

The paper is organized as follows. In Sec.~\ref{II} the basic equations of the theory
 will be given in the adiabatical approximation. In Sec.~\ref{III}, \ref{IV}
  the fluctuations of the Stokes
 parameters will be introduced into consideration, be
 written the respective linearized equations and their solutions. In Sec.~\ref{V}
the comparison of our results with the phenomenological ones will be made. At
last, in Sec.~\ref{VI} the possibilities for the polarization squeezing
observation will be discussed.

\section{The Langevin equations for the two-level lasers with twice
degenerated levels} \label{II} The well known spin-flip VCSEL theory
Ref.~\cite{8} is based on the representation about the semiconductor medium as a
two-level system with a twice degeneration of the levels (see fig.\ 1). One pair
of sub-levels is connected by wave with the $\sigma_+$-polarization and the
other - with the $\sigma_-$ one. The upper "atomic" states $|a_\pm\rangle$ relax
with the rate $\gamma_a$, and the lower - with the rate $\gamma_b$. For the
description of overturns of the electron spins  (the spin-flip) we introduce
into consideration some incoherent processes $|a_+,b_+\rangle\to|a_-,b_-\rangle$
and back  with the respective rates $\gamma_{c}^{(a,b)}$.

We will not mention all the details, how the main equations are produced. They
can be found in the previous work Ref.~\cite{1,2,9}. The main items are as
follows. At first, the equations for the Hermitian operators of the field
amplitudes with the $\sigma_\pm$-polarizations, the populations of the four
sub-levels and the polarizations of the two actual transitions are constructed
on the basis of the full quantum theory for field and matter. The obtained
equations (the Langevin-Heizenberg ones), first, are operator and, second,
contain the operator sources (the inhomogeneous terms) in distinguish from the
respective dynamical theory. Usually for writing the correct correlation
functions one uses so-called Einstein relationships Ref.~\cite{10}.

Next, to make the mathematical situation much simpler, one passes to c-number
representation: all the normally ordered operators can be converted onto
respective c-number functions Ref.~\cite{9,10}. As a result the equations,
called the Langevin's equations,  read:
\BY
     &&\dot{a}_{\pm} =-\kappa a_{\pm}-\(\kappa_a+\i\omega_{p}\)a_{\mp}
     +gP_{\pm},
          \label{2.1}\\
      &&\dot{P}_{\pm}=-\left(\gamma_{\perp}+i\Delta\right)
      P_{\pm}+g\Bigl(N_{a\pm}-N_{b\pm}\Bigr)a_{\pm}
      +F_{P\pm}\(t\),
             \label{2.2}\\
      &&\dot{N}_{a\pm}=\mu_a-\gamma_{a}N_{a\pm}-\gamma_{c}^{(a)}
      \Bigl(N_{a\pm}-N_{a\mp}\Bigr)
              -g\Bigl(a_{\pm}^{\ast} P_{\pm}+a_{\pm}
      P_{\pm}^{\ast}\Bigr)+F_{a\pm}\(t\),
               \label{2.3}\\
      &&\dot{N}_{b\pm}=\mu_b-\gamma_{b}N_{b\pm}-\gamma_{c}^{(b)}
      \Bigl(N_{b\pm}-N_{b\mp}\Bigr)
      +g\Bigl(a_{\pm}^{\ast} P_{\pm}+a_{\pm}
      P_{\pm}^{\ast}\Bigr)+F_{b\pm}\(t\). \L{2.4}
\EY
Here $a_\pm$ are the $\sigma_\pm$ complex amplitudes.  The atomic
 c-number variables are represented by the polarizations $P_\pm$ and
 the populations $N_{a\pm}$ and $N_{b\pm}$. One can see the theory turns out to
 be very complicated mathematically, because generally speaking
 there is a system of  the 12 differential equations relative to the 12 variables.

The frequency coefficients have the physical senses:  $\kappa$ is the spectral
width of the cavity laser mode, $g$ is the atom-field coupling constant,
$\mu_{a,b}$ is the mean rate of the incoherent pump to the upper, lower laser
level, $\gamma_{a,b}$ is the mentioned above constant of decay of the upper,
lower level, $\gamma_c^{(a,b)}$ is the spin-flip rate, $\gamma_\perp$ is the
rate of the transverse atomic relaxation, $\Delta=\omega-\nu$ - detuning of the
laser frequency $\nu$ from the frequency of the laser transition $\omega$, the
coefficients $\kappa_a$ and  $\omega_p$ present the linear dichroizm and linear
birefringence, connected with the optical anisotropy of the semiconductor
crystal.

As for the stochastic sources in the Langevin equations
$F_{b\pm},F_{a\pm},F_{P\pm}$, their properties are given by the following
non-zero correlation functions:
 \BY
     &&\o{F_{a\pm}(t)F_{a\pm}(t^{\prime})} =
     \left [\(\gamma_{a}+\gamma_{c}^{(a)}\)\o N_{a\pm}+\mu_a+\gamma_{c}^{(a)}
     \o N_{a\mp}-g\;\o{\( a_{\pm}^{\ast} P_{\pm}+a_{\pm}
      P_{\pm}^{\ast}\)}-\right.\nn\\
&&\left.-p_a\mu_a/2\right ]\;\delta\!\(t-t^{\prime}\),\L{2.5}\\
     &&\o{ F_{a\pm}(t)F_{a\mp}(t^{\prime})} =-
     \Bigl[\gamma_{c}^{(a)}(\o N_{a\pm}
     +\o N_{a\mp})+p_a\mu_a/2\Bigr]\;
     \delta(t-t^{\prime}),\L{}
                 \\
&&\nn\\ &&\o{F_{b\pm}(t)F_{b\pm}(t^{\prime})} =
     \left [\(\gamma_{b}+\gamma_{c}^{(b)}\)\o N_{b\pm}+\mu_b+\gamma_{c}^{(b)}
     \o N_{b\mp}-g\;\o{\( a_{\pm}^{\ast} P_{\pm}+a_{\pm}
      P_{\pm}^{\ast}\)}-\right.\nn\\
&&\left.-p_b\mu_b/2\right ]\;\delta\!\(t-t^{\prime}\),\\
     &&\o{ F_{b\pm}(t)F_{b\mp}(t^{\prime})} =-
     \Bigl[\gamma_{c}^{(b)}(\o N_{b\pm}
     +\o N_{b\mp})+p_b\mu_b/2\Bigr]\;
     \delta(t-t^{\prime}),
                 \\
&&\nn\\
    &&\o{F_{a\pm}(t)F_{b\pm}(t^{\prime})} =
     g\;\o{\( a_{\pm}^{\ast}
     P_{\pm}+a_{\pm}P_{\pm}^{\ast}\)}\;
     \delta(t-t^{\prime}),
                  \\
&&\nn\\
     &&\o{ F^{\ast}_{P\pm}(t)F_{P\pm}(t^{\prime})}=
     \Bigl[\(2\gamma_{\perp}-\gamma_a-\gamma_{c}^{(a)}\)\o N_{a\pm}+
     \mu_a+\gamma_{c}^{(a)}\o N_{a\mp}\Bigr]\;
     \delta(t-t^{\prime}),
                  \\
    &&\o{F_{P\pm}(t)F_{P\pm}(t^{\prime})}=
     2g\;\o{\( a_{\pm}P_{\pm}\)}\;
     \delta(t-t^{\prime}),
                   \\
   &&\o{F_{P\pm}(t)F_{b\pm}(t^{\prime})}=
     \(\gamma_b+\gamma_{c}^{(b)}\)\;\o {\(P_{\pm}\)}\;
     \delta(t-t^{\prime})\L{2.12}.
\EY
Here the parameters $p_{a,b}$ determine a statistical aspect of the pump: the
Poissonian (quite random) pump takes a place with $p_{a,b}=0$, the
sub-Poissonian (strictly regular) - with $p_{a,b}=1$. Under the calculation of
these formulas it was proposed that distribution of the electrons between
sub-levels turns out to be random even with $p_{a,b}=1$.

The Langevin equations (\ref{2.1})-(\ref{2.4}) with the correlation functions
(\ref{2.5})-(\ref{2.12})
 are suitable, generally speaking, for description
of any two-level laser, because there is no any restriction on the relationship
between the different frequency parameters. Further we will make the area of our
interest narrower adopting our theory for the VCSELs with the equally living
laser levels.

As mentioned in Sec.~\ref{I} the two VCSEL models was developed. In one of them,
which is studied in the San Miguel's group (including the pioneer work
Ref.~\cite{8}), the lifetimes of both the laser levels were equaled. In the
other model
 (the Giacobino's group Ref.~\cite{1,2}) to the contrary the lifetime of
 the lower level was much less  than the lifetime of the upper one.
 As stated above we want to apply our theory for the first model and
 further we will put: $\gamma_a=\gamma_b\equiv\gamma$ and
$\gamma_c^{(a)}=\gamma_c^{(b)}\equiv\gamma_c$. Besides it is suggested the incoherent
pump takes a place only to the upper level: $\mu_a\equiv\mu$ ($\mu_b=0$).

To compare our calculations with the ones, presented in Ref.~\cite{3}, we
introduce new functions $D$ and $d$ instead of the populations $N_{a\pm,b\pm}$:
\BY
&& D=\frac{1}{2}\left [\(N_{a+}+N_{a-}\)-\(N_{b+}+N_{b-}\)\right ]\L{2.13}\\
&& d=\frac{1}{2}\left [\(N_{a+}-N_{a-}\)-\(N_{b+}-N_{b-}\)\right ]\L{2.14}
\EY
and put that the transverse relaxation constant is the highest between the others and we
have a right to apply the adiabatical approximation. In the approximation the equations
read:
 \BY
&&\dot{a}_{\pm} =-\kappa a_{\pm}-\(\kappa_a+\i\omega_{p}\)a_{\mp}
     +c\(1-\i\alpha\)\(D\pm d\)a_\pm +\xi_\pm\L{2.15}\\
 &&\dot{D}=\mu-\gamma D-2cD\(\mid a_+\mid^2+\mid a_{-}\mid^2\)-2cd\(\mid a_
 +\mid^2-\mid a_{-}\mid^2\)+\xi_D\L{2.16}\\
 &&\dot{d}=-\gamma_s d-2cD\(\mid a_+\mid^2-\mid a_{-}\mid^2\)-2cd\(\mid a_+\mid^2+\mid a_{-}\mid^2\)+\xi_d
\L{2.17}\EY
Here
\BE
c=\frac{g^2}{\gamma_\perp(1+\alpha^2)},\qquad\alpha=\frac{\nu-\omega}{\gamma_\perp}
\qquad \gamma_s=\gamma+2\gamma_c.
\EE
The stochastic sources in the equations (\ref{2.15})-(\ref{2.17}) are some linear
combinations of the initial ones. The respective expressions can be found in
 Appendix~\ref{B}.

\section{Fluctuations of the Stokes parameters and linearization of equations}
\label{III} The usual approach to  non-linear equations is to try linearizing them
relative to some small parameter(s). In the cases of statistical theories of laser
systems these small parameters are introduced as additions in the exact solutions to the
stationary solutions.

We will study here only one stationary regime of generation with the linearly
polarized emission. Certainly, everywhere further we must imply the set of the
physical parameters which is able to ensure a stability of this solution
Ref.~\cite{13}.

Following to Ref.~\cite{3,8} the stationary semi-classical solutions can be
written in the form:
 \BE
  a_{\pm
st.}=Q\;e^{-\i\Delta_xt},\qquad \Delta_x=\omega_p+\alpha\kappa_x,\qquad
\kappa_x=\kappa+\kappa_a\L{3.1}
\EE
Here the value $Q$ is expressed via the pump parameter $r=\mu/\mu_{th.}$
($\mu_{th.}$ - the threshold pump rate) as $2Q^2=r-1$.

The equality between the circular components means we have the linearly
polarized along the x-axes non-zero solution. Really, because the Cartesian
field components are expressed via the circular ones in the form:
\BY
a_x=\frac{1}{\sqrt2}(a_++a_-),\qquad a_y=\frac{1}{i\sqrt2}(a_+-a_-)\L{} \EY
$a_x=\sqrt2 Q,\;a_y=0$.

The respective stationary solutions for the active medium are:
 \BY
 &&P_{\pm
st.}=\frac{c}{g}\(1-\i\alpha\)D_{st.}\;Q,\qquad D_{st.}=\frac{\mu}{\gamma+4c\;Q^2},\quad
d_{st.}=0
 \EY
According to (\ref{2.16}) in the absence of the stochastic source  the condition of the
stationary lasing ($\dot D=0$) in the explicit form is: $D_{st}=\kappa_x/c$. This
provides us with a direct connection of the mean radiation power $Q$ with the mean pump
rate of the medium $\mu$:
\BY \frac{\mu}{\gamma+4cQ^2}=\frac{\kappa_x}{c}\L{3.4} \EY
We have discussed the stationary solutions of the problem and now the exact solutions
with taking all the fluctuations into account read:
\BY &&a_\pm=\Bigl(Q+\delta a_\pm(t)\Bigr)\;e^{-\i\Delta_xt},\qquad
D=D_{st.}+\delta D(t),\qquad  d=d_{st}+\delta d(t), \EY
For  the VCSELs  we focus on the  polarization effects. That is why it is
convenient to
 introduce into consideration the Stokes parameters instead of the complex
 field amplitudes. They are expressed via the Cartesian field components as:
\BY
&&S_0=|a_x|^2+|a_y|^2,\;\; S_1=|a_x|^2-|a_y|^2,\;\; S_2=a_x^\ast a_y+a_x a_y^\ast,\;\;
S_3=i(a_x^\ast a_y-a_x a_y^\ast)
\EY
Respectively in our case of the linearly polarized regime their fluctuations are:
 \BY
 && \delta S_1=\sqrt2Q(\delta a_x^*+\delta a_x),\quad\delta S_2= \sqrt2Q(\delta
a_y^*+\delta a_y),\quad \delta S_3= \i\sqrt2Q(\delta a_y^*- \delta a_y) \EY
( $\delta S_0=\delta S_1$) and
 the linearized system of equations reads:
 \BY &&\delta\dot
S_1=\gamma (r-1)\delta D+\xi_{S_1}\L{30} \\ &&\delta\dot D=-\gamma r\delta
D-2\kappa_x\delta S_1+\xi_D\L{31} \EY
and
 \BY &&\delta \dot
S_2=2\kappa_a\delta S_2-2\omega_p\delta S_3-\alpha\gamma (r-1)\delta
d+\xi_{S_2}\\ &&\delta \dot S_3=2\kappa_a\delta S_3+2\omega_p\delta S_2-\gamma
(r-1)\delta d+\xi_{S_3}\\ &&\delta\dot d=-\Gamma_s\delta d+2\kappa_x\delta
S_3+\xi_d \L{32},\qquad \Gamma_s=\gamma_s+\gamma(r-1)=2\gamma_c+\gamma r \L{34}
\EY
One can see there are two independent systems of linear differential equations.
The relation between the stochastic sources
 $\xi_{s_{i=1,2,3}},\;\xi_D,\;\xi_d$ and the initial ones can be found in Appendix~
 \ref{B}.

\section{The spectral densities of fluctuations of the Stokes parameters}
\label{IV} Introducing the Fourie image of the function $F(t)$ as
\BY
 &&
F_{\Omega}=\frac{1}{\sqrt{2\pi}}\int\limits_{-\infty}^{+\infty}F(t)\;e^{\i\Omega
t}\;dt,\qquad
F(t)=\frac{1}{\sqrt{2\pi}}\int\limits_{-\infty}^{+\infty}F_{\Omega}\;e^{-\i\Omega
t}\;d\Omega
\EY
we can rewrite the equations
 (\ref{30})-(\ref{34})  as algebraic ones:
 \BY &&-\i\Omega\;\delta S_{1,\Omega}=\gamma (r-1)\;\delta
D_\Omega+\xi_{S_1;\Omega} \L{37}\\ &&\(-\i\Omega+\gamma r\)\;\delta
D_\Omega=-2\kappa_x\;\delta S_{1,\Omega}+\xi_{D,\Omega}\L{38}\EY
and
 \BY &&-\(\i\Omega+2\kappa_a\)\;\delta
S_{2,\Omega}=-2\omega_p\;\delta S_{3,\Omega}-\alpha\gamma (r-1)\;\delta
d+\xi_{S_2,\Omega}\\ &&-\(\i\Omega+2\kappa_a\)\;\delta
S_{3,\Omega}=2\omega_p\;\delta S_{2,\Omega}-\gamma (r-1)\;\delta
d+\xi_{S_3,\Omega}\\ &&\(-\i\Omega+\Gamma_s\)\;\delta d_\Omega=2\kappa_x\;\delta
S_{3,\Omega}+\xi_{d,\Omega}\EY
Now relatively simply to solve it, and the interesting solutions expressed via
the stochastic sources in the explicit form
 read:
\BY
&&\delta\! S_{1,\Omega}=-\Bigl[\gamma(r-1)\;\xi_{D,\Omega}+(\gamma
r-\i\Omega)\;\xi_{S_1,\Omega}\Bigr]\left [\Omega^2+\i\Omega\gamma
r-2\gamma\kappa_x(r-1)\right ]^{-1}\L{4.7}\\&&\nonumber\\
&& \delta\!
S_{2,\Omega}=1/2\Bigl[\gamma(r-1)\(\omega_p+\alpha\kappa_a+\i\Omega\alpha/2\)
\xi_{d,\Omega}-\(\omega_p\(\Gamma_s-\i\Omega\)+\alpha\gamma\kappa_x(r-1)\)
\xi_{S_3,\Omega}+\Bigr.\nonumber\\
&&\Bigl.+\(-\(\Gamma_s-\i\Omega\)\(\kappa_a+\i\Omega/2\)+\gamma\kappa_x(r-1)\)
\xi_{S_2,\Omega}\Bigr]\times\nonumber\\
&&\Bigl[\(\omega_p^2+\kappa_a^2+\i\Omega\kappa_a-\Omega^2/4\)\(\Gamma_s-\i\Omega\)-\gamma(r-1)\kappa_x\(\kappa_a-\alpha\omega_p+\i\Omega/2\)\Bigr]^{-1}\L{43}\\&&\nonumber\\
&&\delta\!
S_{3,\Omega}=1/2\Bigl[-\gamma(r-1)\(\alpha\omega_p-\kappa_a-\i\Omega/2\)\xi_{d,\Omega}+\omega_p\(\Gamma_s-\i\Omega\)\xi_{S_2,\Omega}-\Bigr.\nonumber\\
&&\Bigl.-\(\Gamma_s-\i\Omega\)\(\kappa_a+\i\Omega/2\)\xi_{S_3,\Omega}\Bigr]
\times\nonumber\\
&&\Bigl[\(\omega_p^2+\kappa_a^2+\i\Omega\kappa_a-\Omega^2/4\)\(\Gamma_s-
\i\Omega\)-\gamma(r-1)\kappa_x\(\kappa_a-\alpha\omega_p+\i\Omega/2\)\Bigr]^{-1}\L{4.9}
\EY
So-called spectral densities of the Stokes parameter fluctuations $(\delta
S_i\;\delta S_k)_\omega$ take an important part under physical discussion. They
are defined under writing the spectral correlation functions:
\BY
\o{\delta S_{i,\Omega}\;\delta S_{k,\Omega^\prime}}=(\delta S_i\;\delta
S_k)_\Omega\;\delta(\Omega+\Omega^\prime)\L{4.10}
\EY
To get the wished spectral densities we must make the last preliminary step and
write the respective spectral densities for the stochastic sources.  According
to Appendix~\ref{B} and (\ref{3.1})-(\ref{3.4}) they are given as:
 \BY
&&(\xi_{S_1}^2)_\Omega =(\xi_{S_2}^2)_\Omega
=(\xi_{S_3}^2)_\Omega=-(\xi_{D}\;\xi_{S_1})_\Omega=(\xi_{d}\;\xi_{S_3})_\Omega=\kappa_x(r^2-1)\;\gamma/c\\
 &&(\xi_{D}^2)_\Omega
=\kappa_x\:r\(r-p/2\)\;\gamma/c\\
 &&(\xi_{d}^2)_\Omega
=\kappa_x\:r\(\gamma_s/\gamma+r-1\)\;\gamma/c
 \EY
We define the non-zero spectral densities of the stochastic sources in the same
way as for the Stokes parameters (\ref{4.10}):
\BY
&&\o{\xi_{S_i,\Omega}\xi_{S_i,\Omega^\prime}}=(\xi_{ S_i}^2)_\Omega
\;\delta(\Omega+\Omega^\prime)\\
&&\o{\xi_{D,d,\Omega}\xi_{D,d,\Omega^\prime}}=(\xi_{D,d}^2)_\Omega
\;\delta(\Omega+\Omega^\prime)\\
&&\o{\xi_{D,\Omega}\xi_{ S_1,\Omega^\prime}}=(\xi_{D}\xi_{ S_1})_\Omega\;\delta(\Omega+\Omega^\prime)\\
&&\o{\xi_{d,\Omega}\xi_{ S_3,\Omega^\prime}}=(\xi_{d}\xi_{ S_3})_\Omega\;
\delta(\Omega+\Omega^\prime)
\EY
Now it is not difficult to get:
\BY
&&(\delta S_1^2)_{\Omega}=n_x\kappa_x \Bigl(2\Omega^2 (r+1)+\gamma^2 r
(p-pr+4)\Bigr)/\lambda_{1,\Omega}\L{4.18}\\
 &&(\delta
S_2^2)_{\Omega}=2n_x\kappa_x\Bigl(\Omega^4/4\;(r+1)+
a_2\Omega^2+b_2\Bigr)/\lambda_{\Omega}\L{4.19}\\
 &&(\delta
S_3^2)_{\Omega}=2n_x\kappa_x\Bigl(\Omega^4/4\;(r+1)+
a_3\Omega^2+b_3\Bigr)/\lambda_{\Omega}\\
 &&(\!\delta\! S_2\delta\!
S_3\!)_{\Omega}=2n_x\kappa_x\gamma(r-1)\Bigl(
a_{23}\Omega^2/4+b_{23}\Bigr)/\lambda_{\Omega}\EY
where the coefficients are defined as:
\BY
&&\nn a_2=(\omega_p^2+\kappa_a^2+\Gamma_s^2/4)(r+1)+\gamma
(r-1)\Bigl(\alpha^2r\Gamma_s/4+(\alpha\omega_p-\kappa_x)(r+1)\Bigr)\\
 &&\nn
b_2=\gamma_s^2(\omega_p^2+\kappa_a^2)(r+1)+\gamma_s\gamma(r-1)\Bigl[r(\omega_p+\alpha\kappa_a)^2+2\kappa(\alpha\omega_p-\kappa_a)(r+1)\Bigr]
+\\ \nn && +\gamma^2(r-1)^2\Bigl[\kappa^2(r+1)(\alpha^2+1)
-(\omega_p+\alpha\kappa_a)^2\Bigr]\\
&&\nn\\
 &&\nn
a_3=(\omega_p^2+\kappa_a^2+\gamma_s^2/4)(r+1)+(\alpha\omega_p-\Gamma_s/2)(r-1)(r+1)\gamma+
\gamma(r-1)r\Gamma_s/4\\
&&\nn
b_3=\Gamma_s^2(r+1)(\kappa_a^2+\omega_p^2)+\gamma(r-1)\Gamma_s\Bigl[r(\alpha^2\omega_p^2-\kappa_a^2)+
2\kappa_a(\alpha\omega_p-\kappa_a)\Bigr]\\
&&\nn\\
 &&\nn
a_{23}=\alpha\Bigl[\Gamma_s+2(r+1)(\kappa-\kappa_a)\Bigr]\\
&&\nn
b_{23}=\gamma_s\Bigl[r(\omega_p+\alpha\kappa_a)(\alpha\omega_p-\kappa_a)+(r+1)\Bigl(\omega_p(\alpha\omega_p-\kappa_a)+\kappa(\omega_p+\alpha\kappa_a)\Bigr)\Bigr]+\\
&&+\gamma(r-1)\Bigl[(\omega_p+\alpha\kappa_a)(\alpha\omega_p-\kappa_a)+(r+1)
(\alpha^2+1)\kappa\omega_p\Bigr]\nn\\
&&\nn\\
 &&\lambda_{1.\Omega}=\(\Omega^2 -
2\gamma\kappa_x (r-1)\)^2+\Omega^2\gamma^2 r^2\nn\\
&&\lambda_{\Omega}=\Omega^2\Bigl[-\Omega^2/4+\kappa_a^2+\omega_p^2-\kappa_a\gamma_s+
\gamma(r-1)(\kappa-\kappa_a)/2\Bigr]^2
+\Bigl[\Omega^2(\kappa_a-\Gamma_s/4)+\Bigr.\nn\\ \nn &&
\Bigl.+\gamma(r-1)\kappa_x(\alpha\omega_p-\kappa_a)+\Gamma_s(\kappa_a^2+\omega_p^2)
\Bigr]^2\nn
 \EY

\section{Comparison with results of the phenomenological calculations}
\label{V} As mentioned above the phenomenological calculations were made in
Ref.~\cite{3} and the respective curves were drown as a result of some numerical
analysis. Here we are going to compare some of these curves with our respective
ones and then  to estimate the phenomenological approach on this basis.

In the fig.\ 2 and 4 are presented the spectral dependences, which we  have
copied simply from the cited Ref.~\cite{3}.  The curves have drown with the
following set of parameters: $\gamma=1 GHz,\;\kappa=300 GHz,\;\omega_p=1
GHz,\;\alpha=-3,\;\gamma_s=100 GHz,\;r=1,04$ in absence of the linear dichroizm
$\kappa_a=0$.

 According Ref.~\cite{3} the curves on Fig.\ 2 are
  the frequency dependences of the
full spectral power of the laser emission (the curve with single maximum)   and
-  of the circularly polarized components (the curve with two maxima). These
match to our $(\delta S_1^2)_{\Omega}$ and $(\delta S_1^2)_{\Omega}+(\delta
S_2^2)_{\Omega}$, which are constructed with help of the formulas
(\ref{4.18})-(\ref{4.19}) and presented in fig.\ 3.

As is seen our  and  obtained phenomenologically curves  turn out to be
qualitatively alike especially as the positions of maxima on the frequency axes
are exactly the same. At the same time one can see the levels of noises in our
approach turn out to be incomparably lower than in the phenomenological one
(please, take into account, the vertical axes on all the pictures  are chosen in
the logarithm scale).

The frequency dependences  for the normalized cross-correlation
\BY
C_{+-}=\frac{(\delta S_1^2)_{\Omega}-(\delta S_2^2)_{\Omega}}{(\delta
S_1^2)_{\Omega}+(\delta S_2^2)_{\Omega}}
 \EY
are presented again respectively for phenomenological (fig.\ 4)
 and our approaches (fig.\ 5). Fixing here
some qualitative (and quantitative - relative to the position of maxima)
similarity, nevertheless one can see some serious differences take a place. For
example,
 the
right side of the graphics after maximum exhibits the absence of correlation in
the phenomenological calculation and the appreciable anti-correlation about
$-1/2$ according to our formulas.

It is easy to understand why the positions of the maxima are the same along the
x-axes in different approaches. It connects only with so-called relaxation
oscillations and that is why this is perfectly independent of the choice of
stochastic sources.

As for quantitative  differences we see the two reasons, why they take a place.
First, under phenomenological introducing the sources to get the respective
correlation functions one must use some additional physical considerations.
Usually, especially under conditions of non-linear field-matter interaction, it
is not a simple problem and requires a special attention. Regrettably this side
of the question has quite fallen out of discussion in Ref.~\cite{3}, and
certainly it provides us with a possibility for doubts relative to  the obtained
correlation functions. At any rate we have to fix our formulas for the
correlation functions of the stochastic sources are quite different.

Second, one of the reasons of our differences is very clear. In Ref.~\cite{3}
the Langevin's equations were produced under introducing the sources into not
the complete system of the dynamical equations but into the simpler system,
where already two variables $D$ and $d$ appear instead of four ones $N_{a\pm}$
and $N_{b\pm}$. We remember use of the simpler system of equations
 turned out to be possible
in the dynamical theory of the VCSELs with the equaled relaxation constants of
both the levels. But in the statistical theory it leads to the losses of the
important sources, which depend on the populations of the sub-levels $N_{a\pm}$
and $N_{b\pm}$. We think it is main reason why the correlation functions in
Ref.~\cite{3} are proportional only to the population differences of the kind of
$N_{a\pm}-N_{b\pm}$, what can not be correct.

\section{Polarization squeezing in the VCSEL emission}\label{VI}
In Ref.~\cite{2} we have discussed in details how to observe the Stokes
parameter fluctuations in experiments with two photodetectors. Choosing the
available geometry of the experiment we are able to select the wished signal.
The most important cases read:
 \BY
&&(\delta i_{-}^2)_\Omega/\langle i_+\rangle=1+2\kappa/n_x(\delta
S_{1}^2)_\Omega,\qquad\varphi=0;\L{6.1}\\
   \quad&&(\delta i_{-}^2)_\Omega/\langle i_+\rangle=1+2\kappa/n_x(\delta S_{2}^2)_\Omega,\qquad\varphi=\pi/4,\;\theta=0;\\
\L{61}\quad&&(\delta i_{-}^2)_\Omega/\langle i_+\rangle=1+2\kappa/n_x(\delta
S_{3}^2)_\Omega,\qquad\varphi=\pi/4,\;\theta=\pi/2;\L{62} \EY
and
 \BY
&&\varphi=\pi/4,\;\theta=\pi/4:\nn\\ &&(\delta i_{-}^2)_\Omega/\langle
i_+\rangle=1+\kappa/n_x\Bigl[(\delta S_{2}^2)_\Omega +(\delta
S_{3}^2)_\Omega+2(\delta\! S_{2}\delta\! S_{3})_\Omega\Bigr],\L{63}\\
&&\varphi=\pi/8,\;\theta=0:\nn\\&&(\delta i_{-}^2)_\Omega/\langle
i_+\rangle=1+\kappa/n_x\Bigl[(\delta S_{1}^2)_\Omega +(\delta S_{2}^2)_\Omega+ 2
(\delta\! S_{1}\delta\! S_{2})_\Omega\Bigr],\L{64}\\
&&\varphi=\pi/8,\;\theta=\pi/2:\nn\\&&(\delta i_{-}^2)_\Omega/\langle
i_+\rangle=1+\kappa/n_x\Bigl[(\delta S_{1}^2)_\Omega +(\delta S_{3}^2)_\Omega+
2(\delta\! S_{1}\delta\! S_{3})_\Omega\Bigr].\L{65}
\EY
Here $\varphi$ is the angle between the direction of the  linear polarization of
the VCSEL emission and the polarization beam-splitter axis,
 $\theta$ is additional phase shift, introducing by the phase plates between
 the orthogonal field components.

Our theory elaborated in the previous sections gives the possibilities to study
any signals. But here we will discuss only the first of them (\ref{6.1}),
connected with the polarization squeezing and that is why having the principal
character for quantum optics. As for the others to our mind it is interesting to
consider them together with the  respective experimental date for comparison.

In semiconductor lasers (including the VCSELs) it is easy achieved the
regularity in the pump, that, as we remember, leads    to the essential
intensity noise  reduction
 below quantum limit for
 two level lasers without any degeneration. The degeneration of the laser levels
 introduces to system
some additional random process, namely electrons under pumping are distributed
between the sub-levels quite accidentally. This turned out to be inessential for
the lasers with the shortly living lower level, because there  in the linearly
polarized emission only the full population of both the upper laser sub-levels
$N_{a+}+N_{a-}$ plays the role for squeezing. And just it does not fluctuate
under the regular pump.

At the same time in our case the situation appears more complicated for
understanding  because now the value $(N_{a+}+N_{a-}-N_{b+}-N_{b-})$ produces
squeezing.

To make some conclusion it is enough to watch only over the point $\Omega=0$ in
the spectrum (\ref{6.1}). It means we want to watch over  a depth of the noise
reduction below the quantum limit. Putting $p=1$ (the regular pump) and
$\kappa_a=0$ (without the dichroizm) we obtain that
\BY &&(\delta i_-^2)_{\Omega=0}/\langle
i_+\rangle=1+\frac{1}{2}\;\frac{5-r}{r-1}\;\frac{r}{r-1}\;
\stackrel{r\gg1}\longrightarrow \frac{1}{2}\L{6.7} \EY
Comparing it with  the respective formula for the laser without the level
degeneration \cite{5}:
\BY &&(\delta i_-^2)_{\Omega=0}/\langle i_+\rangle=1+\frac{1}{2}\;\frac{5-r}{r-1}\;
\stackrel{r\gg1}\longrightarrow \frac{1}{2}\L{6.8} \EY
one can see there is an additional factor in (\ref{6.7}) equaled to $r/(r-1)$.
It is clear this plays an important  role for a small amount of the pump
parameter $r$ (especially just near the threshold with $r-1\ll1$). At the same
time, squeezing is able to appear only with the high enough pump parameter,
namely with $r>5$. Then  the additional factor is already about one and hence
turns out to be quite inessential. So our general conclusion is the degeneration
of the laser levels does not leads to some additional difficulties in the
production of squeezing.

Also it is interesting to compare two kind of the VCSELs with the shortly living
lower level (the case in Ref.~\cite{2}) and with the equally living levels. The
respective formula in Ref.~\cite{2} is:
\BY
&&(\delta i_-^2)_{\Omega=0}/\langle
i_+\rangle=1+\frac{3-r}{r-1}\;\frac{r}{r-1}\stackrel{r\gg1}\longrightarrow0\L{6.9}
\EY
For example, if we choose $r=6$, then according to the last formula we have the
noise level is 0,28 of the quantum limit. At the same time for our case here it
is only 0,84. One can see the reduction of noises below the quantum limit is
more effective for the VCSEL with the shortly living lower laser level.

\section{Conclusion}
To conclude we would like to say once more which concrete targets have been
achieved in this work. First of all, the full quantum-statistical theory of the
VCSEL with the equally living laser levels has been built. Thereby now we have a
possibility to consider the problem on the same level of understanding as for
the VCSEL with the shortly living lower level.

On this base  the spectral densities of the Stokes parameter fluctuations have
been written under taking into account
 the optical anisotropy of the semi-conductor crystal (the linear dichroizm and
 the linear birefringence) and the spin-flip.

We have been discussed the experimental situation in which it is possible to
observe any spectral densities of the Stokes parameter fluctuations and also
their correlations. Under discussion our main attention was devoted to the
problem of polarization squeezing and the role of the degeneration of the laser
levels. We concluded that the role of the random distribution of the electrons
between the sub-levels under pumping is inessential for the shot noise
reduction.

\begin{acknowledgments}
This work was performed within the Franco-Russian cooperation program
``Lasers and Advanced Optical Information Technologies'' with financial
support from the following organizations: INTAS (grant INTAS-01-2097),
RFBR (grant 03-02-16035), Minvuz of Russia (grant E 02-3.2-239), and by
the Russian program ``Universities of Russia'' (grant ur.01.01.041).
\end{acknowledgments}



\appendix
\section{Stochastic sources in the linearized Langevin equations}\label{B}
The basic equations of the theory (\ref{2.1})-(\ref{2.4}) have the stochastic
sources
 as inhomogeneous terms. The properties of them are specified by the non-zero
 correlation functions (\ref{2.5})-(\ref{2.12}). The case of our interest is
 $\mu_a\equiv\mu,\quad\mu_b=0$ and $\gamma_a=\gamma_b\equiv\gamma$ and
$\gamma_{c}^{(a)}=\gamma_{c}^{(b)}\equiv\gamma_c$. We can rewrite the equations
in the form which they is given in Ref.~\cite{3} in:
 \BY &&\dot{a}_{\pm}
=-\kappa a_{\pm}-\(\kappa_a+\i\omega_{p}\)a_{\mp}
     +gP_{\pm},
          \label{cnumber1}\\
      &&\dot{P}_{\pm}=-\left(\gamma_{\perp}+i\Delta\right)
      P_{\pm}+g\Bigl(D\pm d\Bigr)a_{\pm}
      +F_{P\pm}\(t\),
             \label{cnumber2}\\
      &&\dot{D}=\mu-\gamma D-g\Bigl(a_{+}^{\ast} P_{+}+a_{+}
      P_{+}^{\ast}+a_{-}^{\ast} P_{-}+a_{-}
      P_{-}^{\ast}\Bigr)+F_{D}\(t\),
             \\
      &&\dot{d}=-\gamma_{s}d
      -g\Bigl(a_{+}^{\ast} P_{+}+a_{+}
      P_{+}^{\ast}-a_{-}^{\ast} P_{-}-a_{-}
      P_{-}^{\ast}\Bigr)+F_{d}\(t\),\qquad \gamma_s=\gamma+2\gamma_c.
\EY
Here instead of the populations $N_{a\pm}$ and $N_{b\pm}$  the new variables are
introduced:
\BY &&D=\frac{1}{2}\Bigl[\(N_{a+}+N_{a-}\)-\(N_{b+}+N_{b-}\)\Bigr]\\
&& d=\frac{1}{2}\Bigl[\(N_{a+}-N_{a-}\)-\(N_{b+}-N_{b-}\)\Bigr], \EY
Respectively instead of the initial sources $F_{a\pm}$ and $F_{b\pm}$ the new
ones are created:
\BY
&&F_D=\frac{1}{2}\(F_{a+}+F_{a-}-F_{b+}-F_{b-}\)\\
&&F_d=\frac{1}{2}\(F_{a+}-F_{a-}-F_{b+}+F_{b-}\)
 \EY
The correspondent non-zero correlation functions read:
 \BY
&&\o{F_{D}(t)F_{D}(t^{\prime})}=\frac{1}{4}\left [\gamma\(\o N_{a+}+\o N_{a-}+\o
N_{b+}+\o N_{b-}\)+2\mu\(1-p\)\right.\nn\\ &&\left.-4g\(\o{a_{+}^{\ast}
P_{+}}+\o{a_{+}P_{+}^{\ast}}+\o{a_{-}^{\ast}
P_{-}}+\o{a_{-}P_{-}^{\ast}}\)\right ]\;\delta(t-t^{\prime}). \\
&&\o{F_{d}(t)F_{d}(t^{\prime})}=\frac{1}{4}\left [\(\gamma+4\gamma_c\)\(\o
N_{a+}+\o N_{a-}+\o N_{b+}+\o N_{b-}\)+2\mu-\right.\nn\\
&&\left.-4g\(\o{a_{+}^{\ast} P_{+}}+\o{a_{+}P_{+}^{\ast}}+\o{a_{-}^{\ast}
P_{-}}+\o{a_{-}P_{-}^{\ast}}\)\right ]\;\delta(t-t^{\prime}).
\\
&&\o{F_{D}(t)F_{d}(t^{\prime})}=\o{F_{d}(t)F_{D}(t^{\prime})}=\frac{1}{4}\left
[\gamma\(\o N_{a+}-\o N_{a-}+\o N_{b+}-\o N_{b-}\)-\right.\nn\\
&&\left.-4g\(\o{a_{+}^{\ast} P_{+}}+\o{a_{+}P_{+}^{\ast}}-\o{a_{-}^{\ast}
P_{-}}-\o{a_{-}P_{-}^{\ast}}\)\right ]\;\delta(t-t^{\prime}).\\
&&\o{F_{D}(t)F_{P\pm}(t^{\prime})}=-\frac{1}{2}\(\gamma+\gamma_c\) \o
P_\pm\;\delta(t-t^{\prime}).\\
&&\o{F_{d}(t)F_{P_\pm}(t^{\prime})}=\mp\frac{1}{2}\(\gamma+\gamma_c\) \o
P_\pm\;\delta(t-t^{\prime}).\\ &&\o{ F^{\ast}_{P\pm}(t)F_{P\pm}(t^{\prime})}=
     \Bigl[\(2\gamma_{\perp}-\gamma\)\o
     N_{a\pm}-\gamma_c\(\o
     N_{a\pm}-\o N_{a\mp}\)+\mu\Bigr]\;
     \delta(t-t^{\prime}),
                  \\
    &&\o{F_{P\pm}(t)F_{P\pm}(t^{\prime})}=
     2g\;\o{\( a_{\pm}P_{\pm}\)}\;
     \delta(t-t^{\prime}),
\EY
Under the adiabatical approximation $\dot P_\pm=0$, and we can get:
\BY &&P_\pm=\frac{1}{\gamma_\perp+\i\Delta}\left [g\(D\pm
d\)a_\pm+F_{P\pm}\right ] \EY
Taking this into account  we can write our basic equations in the adiabatical
approximation (\ref{2.15})-(\ref{2.17}). There the sources are:
\BY &&\xi_\pm=\frac{c}{g}\(1-\i\alpha\)F_{P\pm}
       \\
&&\xi_D=F_D-\(a_+^\ast \xi_++a_-^\ast \xi_-+a_+ \xi_+^\ast+a_- \xi_-^\ast\)\\
&&\xi_d=F_d-\(a_+^\ast \xi_+-a_-^\ast \xi_-+a_+ \xi_+^\ast-a_- \xi_-^\ast\)
 \EY
The linearization of the equations relative to the fluctuations near the
stationary semi-classical solutions leads to the equations(\ref{30})-(\ref{34})
which the fluctuations of the Stokes parameters are introduced in and
correspondently  the new sources read:
\BY &&\xi_{S_1}=\sqrt2Q(\xi_xe^{\i\Delta_x t}+\xi_x^\ast e^{-\i\Delta_x t})\\
&&\xi_{S_2}=\sqrt2 Q(\xi_ye^{\i\Delta_x t}+\xi_y^\ast e^{-\i\Delta_x t})\\
&&\xi_{S_3}=\i \sqrt2 Q(\xi_y^\ast e^{-\i\Delta_x t}-\xi_y e^{\i\Delta_x t})\\
&&\xi_D=F_D-\sqrt2 Q(\xi_x+\xi_x^\ast )\\ &&\xi_d=F_d-\i \sqrt2
Q(\xi_y-\xi_y^\ast ) \EY
where
\BY
&&\xi_x=\frac{1}{\sqrt2}\(\xi_++\xi_-\)=\frac{c}{\sqrt2g}\(1-\i\alpha\)\(F_{P+}+F_{P-}\)\\
&&\xi_y=\frac{1}{\i\sqrt2}\(\xi_+-\xi_-\)=\frac{c}{\i\sqrt2g}\(1-\i\alpha\)\(F_{P+}-F_{P-}\)
 \EY

 \newpage

 \begin{figure}[t]
  \includegraphics[width=120mm]{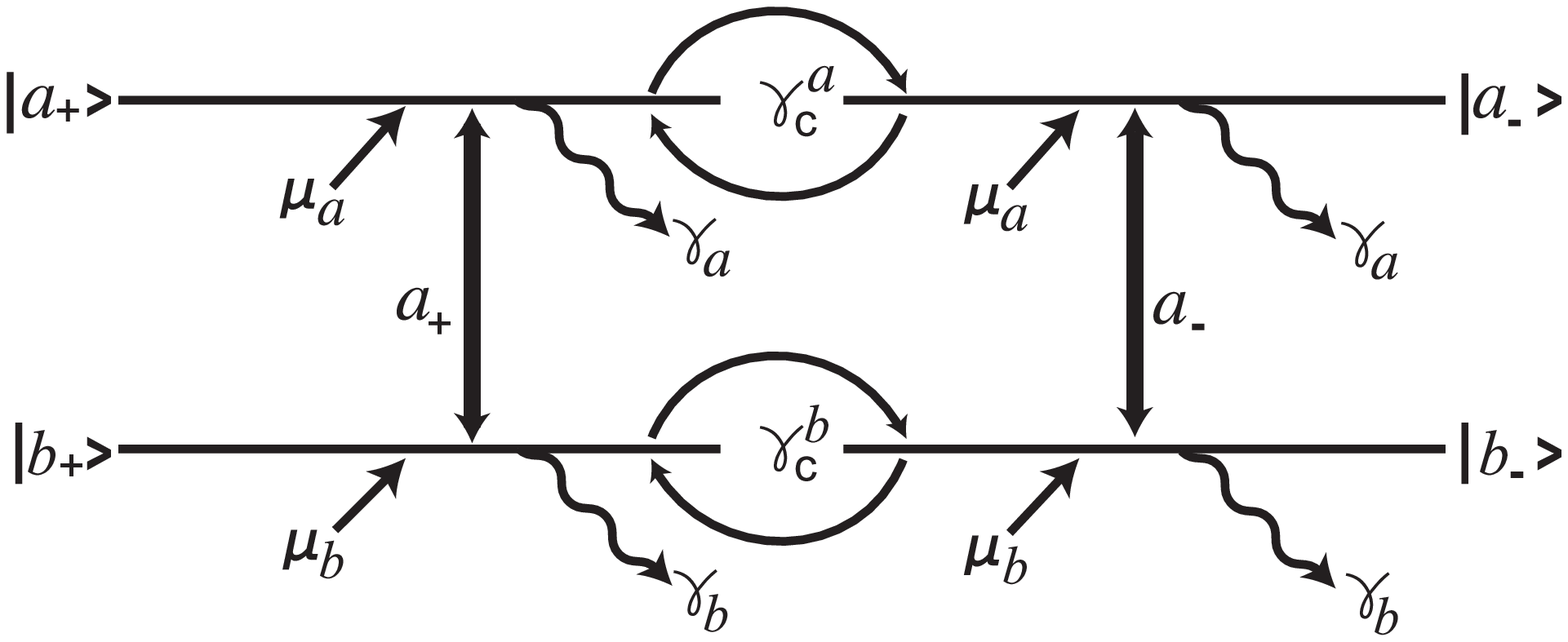}
 \caption{Four-level configuration of the VCSEL medium}
 \label{fig1}
 \end{figure}

 \begin{figure}[t]
  \includegraphics[width=120mm]{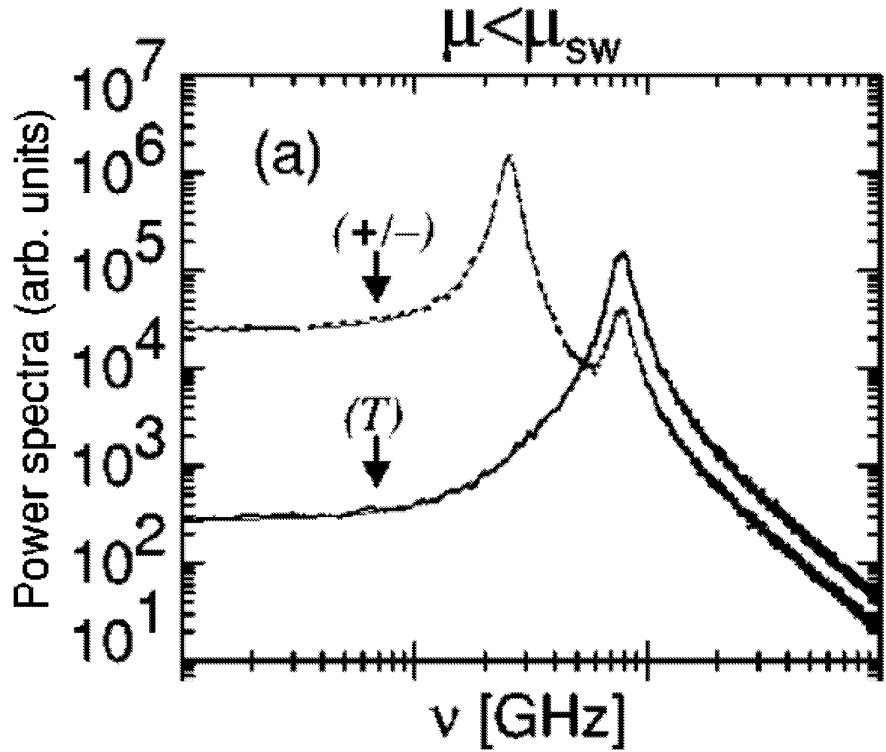}
 \caption{ The full spectral power (fragment of the figure 1 \cite{3})with
 the following set of parameters:
 $\kappa=300 GHz,\;\omega_p=1 GHz,\;; \alpha=-3,\; \gamma=1 GHz,\;
  \gamma_s=100 GHz$ and $\kappa_a=0$, $r=1.04$}
 \label{fig2}
 \end{figure}

 \begin{figure}[t]
  \includegraphics[width=120mm]{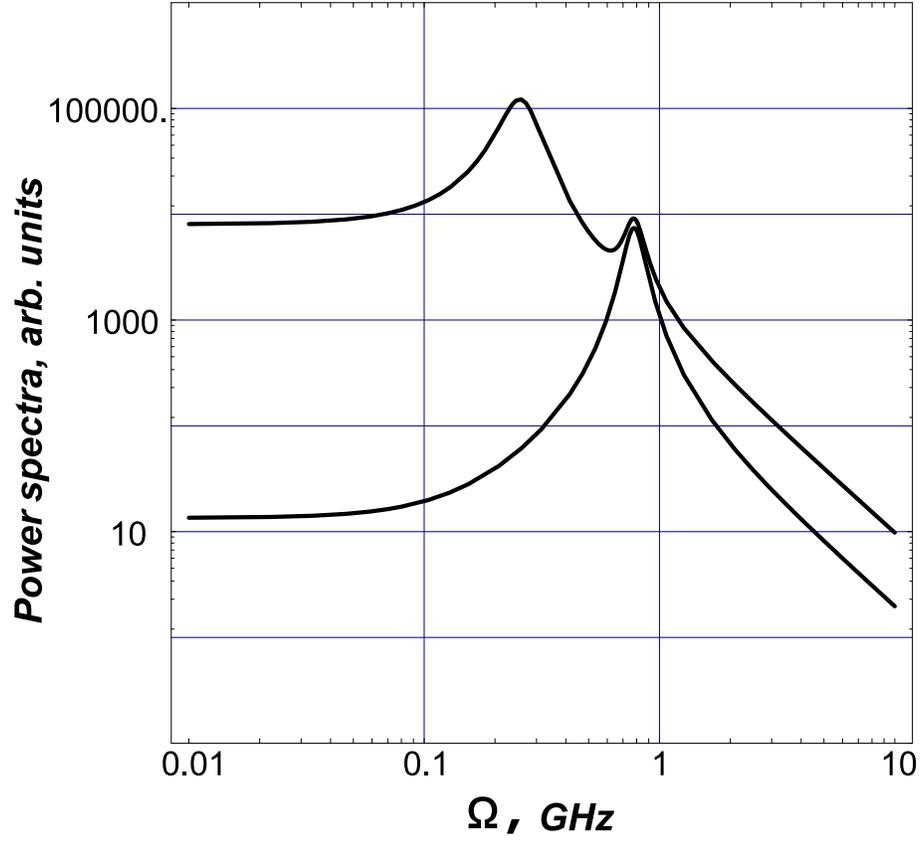}
 \caption{ The full spectral power (numerical calculation in our approach) with
 the following set of parameters:
 $\kappa=300 GHz,\;\omega_p=1 GHz,\;; \alpha=-3,\; \gamma=1 GHz,\;
  \gamma_s=100 GHz$ and $\kappa_a=0$, $r=1.04$}
 \label{fig3}
 \end{figure}

 \begin{figure}[t]
  \includegraphics[width=120mm]{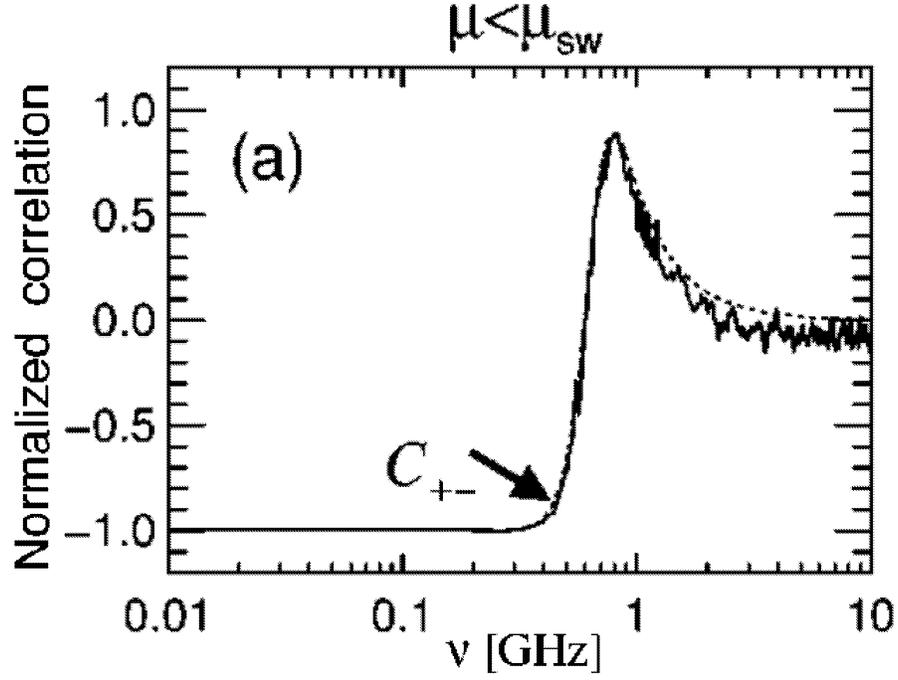}
\caption{The frequency dependence of the correlations between the circularly polarized
field components (fragment of the figure 1 \cite{3})) with
 the following set of parameters:
 $\kappa=300 GHz,\;\omega_p=1 GHz,\;; \alpha=-3,\; \gamma=1 GHz,\;
  \gamma_s=100 GHz$ and $\kappa_a=0$, $r=1.04$}
 \label{fig4}
 \end{figure}

 \begin{figure}[t]
  \includegraphics[width=120mm]{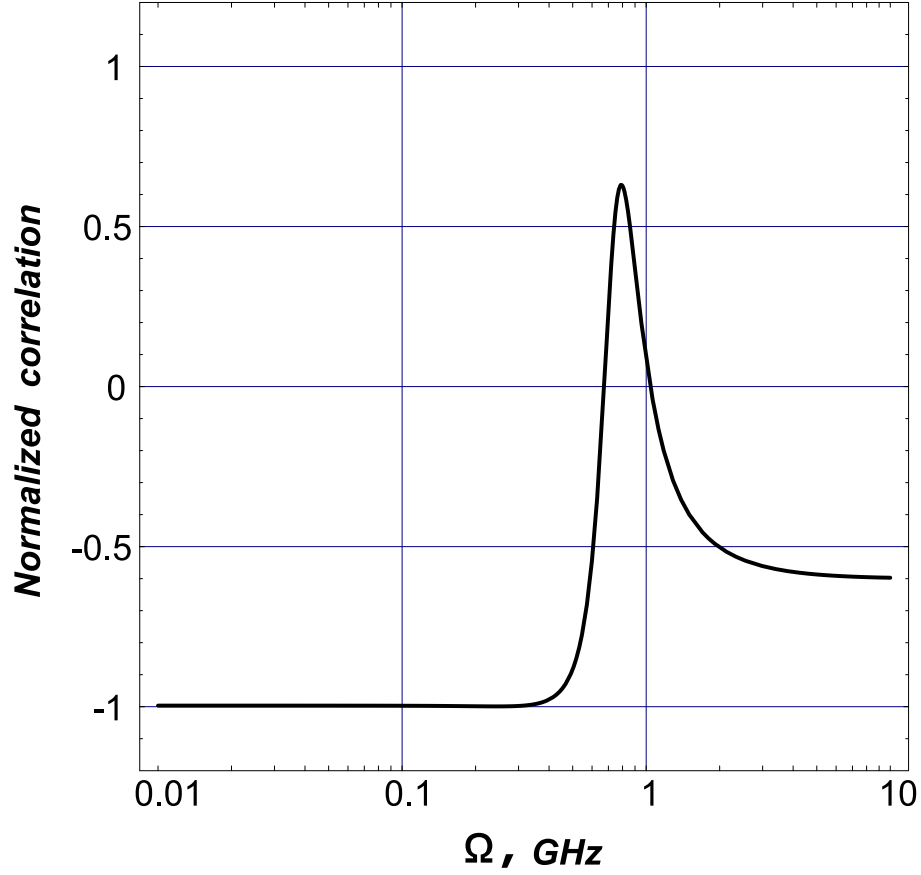}
 \caption{ The frequency dependence of the correlations between the circularly polarized
field components (numerical calculation in our approach) with
 the following set of parameters:
 $\kappa=300 GHz,\;\omega_p=1 GHz,\;; \alpha=-3,\; \gamma=1 GHz,\;
  \gamma_s=100 GHz$ and $\kappa_a=0$, $r=1.04$}
 \label{fig5}
 \end{figure}

\end{document}